\documentclass[11pt]{article}
\usepackage{color}
\usepackage{amsmath,amssymb}
\usepackage{graphicx} 
\usepackage{bbm}
\usepackage{longtable}
\usepackage[small,loose]{subfigure}
\usepackage[small]{caption2} 
\usepackage{fleqn} 
\usepackage{cite}
\usepackage{pifont}

\advance \headheight by 3.0truept       
\setcaptionwidth{0.85\textwidth}

\newcommand{\E}[1]{\ensuremath{\mathrm{E}_{#1}}} 

\newcommand{\SO}[1]{\ensuremath{\mathrm{SO}(#1)}}
\newcommand{\SU}[1]{\ensuremath{\mathrm{SU}(#1)}}
\newcommand{\U}[1]{\ensuremath{\mathrm{U}(#1)}}
\newcommand{\Z}[1]{\ensuremath{\mathbbm{Z}_{#1}}} 

\allowdisplaybreaks

\begin{document}

\date{}
\title{
\begin{flushright}
\normalsize{DESY-09-213}\\
\end{flushright}
\vskip 2cm
{\bf\huge The NMSSM and String Theory}\\[0.8cm]}

\author{{\bf\normalsize
Oleg~Lebedev and
Sa\'ul~Ramos-S\'anchez}\\[1cm]
{\it\normalsize
DESY Theory Group}\\ 
{\it\normalsize Notkestrasse 85, D-22603 Hamburg, Germany
}
}

\maketitle 

\thispagestyle{empty}

\vskip 1cm
\begin{abstract}
We study the possibility of constructing the NMSSM from the heterotic
string.  String derived NMSSMs are much more rare than MSSMs due to
the extra requirement that there exist a light singlet which couples
to the Higgs pairs. They share the common feature that the singlet
self--interactions are typically suppressed, leading to either the
``decoupling'' or the Peccei-Quinn limit of the NMSSM. In the
latter case, the spectrum contains a light pseudoscalar which may be
relevant to the MSSM fine-tuning problem. 
We provide a \Z6 heterotic orbifold   example of the
NMSSM with approximate Peccei-Quinn symmetry, whose origin 
lies in  the string
selection rules combined with our  choice of the vacuum configuration. 
\end{abstract}
\clearpage

\newpage

\section{Introduction}

The next--to--minimal supersymmetric Standard Model (NMSSM) is a minimal extension of the MSSM which
includes a Standard Model (SM) singlet $S$ (for recent reviews, see
\cite{Ellwanger:2009dp,Maniatis:2009re}).  It has certain advantages over the MSSM in that (i) it can
provide a solution to the $\mu$--problem, (ii) it requires milder fine-tuning to accommodate the LEP Higgs
bound. In the NMSSM, the SM--like Higgs boson $h$ can have unusual decay channels. If there is a light
pseudoscalar $a$, $h$ can predominantly decay into pairs of $a$'s, which subsequently decay into taus or
light quarks~\cite{Dermisek:2005ar}. For such final states, the LEP bound on the Higgs mass relaxes and
can in some cases be as low as 90 GeV. As a result, the superpartners are not required to be very heavy
for the Higgs mass to comply with the bound, and the fine-tuning problem of the MSSM can be avoided.

Motivated by these considerations, we undertake a search for string--derived NMSSMs. Recently, a number
of different approaches have yielded examples of models with the exact MSSM spectrum. These include
heterotic \Z6~\cite{Buchmuller:2005jr,Lebedev:2006kn}, \Z{12}~\cite{Kim:2006hv},
\Z2$\times$\Z2~\cite{Blaszczyk:2009in} orbifolds as well as smooth Calabi--Yau
compactifications~\cite{Bouchard:2005ag,Braun:2005nv,Anderson:2009mh} of the heterotic
string.\footnote{This lists extends further if one allows for vector--like
  exotics~\cite{Ibanez:1987sn,Font:1988mm,Casas:1988se,Dijkstra:2004cc,Blumenhagen:2006ux,Faraggi:2006bc,Gmeiner:2008xq}.
}  In fact, it has been shown in Refs.~\cite{Lebedev:2006kn,Lebedev:2008un} that there is a ``fertile
patch'' in the heterotic landscape, where more than 0.1\% of all inequivalent models have the MSSM
spectrum. This result is based on the concept of ``local GUTs''~\cite{Buchmuller:2005jr}, which has been
prompted by the orbifold GUT interpretation of the heterotic string models
\cite{Kobayashi:2004ud,Forste:2004ie,Buchmuller:2004hv}.  The main idea is that twisted matter comes from
points in the compact space with \SO{10} or \E6 GUT symmetry and thus forms complete representations of
the corresponding gauge group, while the (untwisted) gauge fields only respect the Standard Model
symmetry in 4 dimensions.  Many of the resulting models have a number of phenomenologically attractive
features including R--parity \cite{Lebedev:2007hv}, the neutrino seesaw \cite{Buchmuller:2007zd} and
preference for the TeV scale soft SUSY breaking masses \cite{Lebedev:2006tr}.

In what follows, we explore the ``fertile patch'' (and beyond) of the heterotic minilandscape
to identify the NMSSM candidates\footnote{Some studies of the singlet extensions of the Pati-Salam
model have been performed in \cite{Chemtob:2009ie}.}. Common features of these models are discussed
in Section 2, while a specific example is presented in Section 3.

\section{Generalities}

The  relevant superpotential is given by 
\begin{equation}
\label{nmssm}
W= \lambda S H_u H_d + {1\over 3}\kappa S^3 \;,  
\end{equation}
which corresponds to the ``\Z3--symmetric'' NMSSM\footnote{This
symmetry is only approximate. For example, it can be broken by a
(small) supersymmetric mass term for $S$. }.  Here we are considering
$S$ to be a ``massless'' at the string level singlet in the sense that
its supersymmetric mass term is well below the electroweak scale.
Furthermore, we are assuming that supersymmetry is not broken by the
F--term of $S$ (in the limit $\langle S \rangle \rightarrow 0$) and
thus the ``tadpole'' term linear in $S$ is also negligible.  The
corresponding soft supersymmetry breaking terms are
\begin{equation}
-{\cal L}_{\rm soft}= m_S^2 \vert S\vert^2 + \Bigl(\lambda A_\lambda ~ S H_u H_d + 
{1\over 3}\kappa A_\kappa  ~S^3  + {\rm h.c.}   \Bigr)\;, 
\end{equation}
where we have omitted $S$-independent  terms. In what follows, we will assume
all soft terms to be of the same order of magnitude (EW size). We will also
omit the CP-- phases which are strongly constrained by electric dipole moments 
(see, e.g. \cite{Abel:2001vy}).

In general, $\lambda $ and $\kappa$ are the effective couplings,  
\begin{equation}
\lambda = {\rm const} + \langle s_{a_1}...s_{a_n}   \rangle ~,~ 
\kappa =  \langle s_{b_1}...s_{b_n}   \rangle \;, \label{c}  
\end{equation}
where $s_i$ are Standard Model singlets which get non-negligible  VEVs in Planck units and
the ``const'' indicates a direct trilinear coupling. 
The Standard Model singlet $S$ comes originally from the E$_8 \times $E$_8  $ sector 
of the theory and therefore carries charges under some of the gauge groups.
The $S^3$ interaction violates these symmetries and 
only after gauge symmetries  get broken spontaneously, is this effective interaction allowed.
As a result, $\kappa$ is suppressed by  the SM singlet VEVs 
$ \langle s_{b_1}...s_{b_n}   \rangle. $\footnote{There is a caveat here: the massless singlet
can be a linear combination of the type $N=(a S_1 +b S_2)/\sqrt{a^2 +b^2}$,
where $a,b$ are proportional to some  VEVs. In this case, 
unsuppressed $N^3$--interaction 
 may be allowed, although it is singular in the limit of vanishing VEVs. We find that in practice it does not happen since the string selection rules are very constraining. } (If $S$ comes from the gravitational sector,
it is neutral under gauge symmetries but its interactions are suppressed). 
In contrast, a coupling among three
different fields is allowed already at the trilinear level, hence the ``const'' term
in Eq.(\ref{c}).

The size of $\langle s_i \rangle $ is model-dependent. Some SM singlets are 
required to attain VEVs by supersymmetry. In particular, the presence of an 
anomalous U(1) symmetry   induces  the Fayet-Iliopoulos (FI)  term \cite{Dine:1987xk}   
\begin{equation}
D_{\rm FI}={ g M_{\rm Pl}^2 \over 192 \pi^2}~ {\rm Tr} ~\U1_{\rm anom} + \sum_i q^i 
\vert s_i \vert^2 \;,  
\end{equation}
which must vanish in a supersymmetric configuration. 
Here $q^i$ are the anomalous \U1 charges of $s_i$ and $g$ is the gauge coupling.
Since 
${\rm Tr}~ \U1_{\rm anom} \not= 0$, some singlets develop VEVs somewhat below
the Planck  scale. This sets the scale for other singlet VEVs as well and
one generally expects $\langle s_i \rangle $ to be in the range ${\cal O}(10^{-1})-
{\cal O}(10^{-2})$ in Planck units, although ${\cal O}(1)$ VEVs  are also possible.

Therefore, typically
\begin{equation} 
\kappa \ll 1 \;,
\end{equation}
while $\lambda$ can be order one. If  the ``const'' term
in Eq.(\ref{c}) vanishes due to string selection rules, then $\lambda$ is also
suppressed. We thus are led to two distinct versions of the NMSSM:
the ``decoupling'' ($\lambda,\kappa \ll 1$)   and the Peccei-Quinn scenarios
($\kappa \ll 1$). Let us consider these limits in more detail, following Ref.~\cite{Ellwanger:2009dp}.

1) {\it Decoupling limit}.  For $\lambda,\kappa \ll 1$, the singlet essentially
decouples and the NMSSM degenerates into a version of the MSSM, albeit with
modifications in the neutralino sector.
The potential for  (the real part of) the scalar component of
$S$, denoted by $s$, is given by
\begin{equation}    
V(s) \sim m_S^2 s^2  + {2\over 3} \kappa A_\kappa s^3 + \kappa^2 s^4 \;.
\end{equation}
For $A_\kappa^2 \geq 8 m_S^2$, there is a local minimum at
\begin{equation} 
s \simeq {1\over 4 \kappa} \Bigl(   - A_\kappa + \sqrt{A_\kappa^2 -8 m_S^2}  ~   \Bigr)\;.
\end{equation}
Since the chargino mass bound requires $\lambda s \sim~ $EW, we have
\begin{equation}
 s \sim {{\rm EW} \over \kappa} \sim  {{\rm EW} \over \lambda}
\end{equation}
for the soft terms of the electroweak size. This defines the decoupling limit. 
The difference from the MSSM resides in the neutralino sector: the fermionic
component of $S$ has mass $2\kappa s$ and can be the LSP. The NLSP decays are 
then suppressed by the small coupling $\lambda$ leading to its   
long lifetime with  characteristic signatures such as   displaced  vertices \cite{Ellwanger:1997jj}.

2) {\it Peccei--Quinn limit}. For $\kappa \ll 1$ \cite{Miller:2003ay}, 
the model possesses an approximate 
 Peccei--Quinn symmetry $H_{u,d} \rightarrow e^{i \alpha} H_{u,d} $, $S \rightarrow
e^{-2 i \alpha} S$. Spontaneous breaking of this symmetry generates a pseudo--Goldstone
boson (axion). The composition of this state is given by
\begin{eqnarray}
&& A_{\rm PQ} = {1\over \sqrt{v^2 \sin^2 2\beta +4 s^2}} ~(v \sin 2\beta~A   -2 s~ S_I)\;,
\nonumber\\
&& A= \cos\beta ~H_{uI}+\sin\beta~H_{d I} \;,
\end{eqnarray}
where $S_{I}, H_{uI},H_{dI}$ are defined by $f_I \equiv \sqrt{2}~ {\rm Im}(f-\langle f \rangle) 
$. As usual, $\tan\beta=v_u/v_d$ and $v=\sqrt{v_u^2 +v_d^2}=174$ GeV. 

The presence of a light axion--like state  can be relevant to the MSSM fine-tuning problem
\cite{Dermisek:2005ar}.
Typically, $s\gg v \sin 2\beta$, so that the axion is predominantly an EW singlet.
Its couplings to quarks and gauge bosons are suppressed, but the coupling 
to the Higgs bosons is significant. Thus the SM-like Higgs $h$ can decay into pairs
of $ A_{\rm PQ}$ which would subsequently decay into 4 fermions. If $m_{A_{PQ}}< 2 m_b$,
the dominant decay channel would be $h\rightarrow 2  A_{\rm PQ} \rightarrow 4 \tau ~(4 q)$,
with $q$ being light quarks\footnote{There are further constraints on this scenario from
meson decays \cite{Dermisek:2006py,Domingo:2008rr}.}.
Under these conditions, the LEP bound on the Higgs mass relaxes to about 105 GeV
for the final state taus    and  90 GeV for the final state light quarks \cite{newbound}.
This ameliorates the MSSM fine-tuning problem since the  superpartners are not  required
to be very heavy  to accommodate the LEP Higgs  bound.

Let us conclude this section by noting that the above symmetries may appear puzzling
from the low energy perspective. They are consequences of the stringy   UV completion
of these effective theories. For example, the absence or suppression of the $S^3$ term
has to do with the fact that $S$ is charged under additional gauge symmetries. Similarly,
the absence  of the direct $\mu$-term is a result of the string selection rules and
our choice of the vacuum state. It is  interesting that  these string constructions   
favor  certain versions of the NMSSM.

\section{Search for the NMSSM}

The \Z6-II heterotic orbifold is known to yield many examples of models with the MSSM spectrum.
Particularly favorable  are the gauge embeddings which produce local GUTs like \SO{10} or \E6
at some fixed points. Properties of these models are summarized in 
Refs.~\cite{Lebedev:2006kn,Lebedev:2008un}.
Clearly, to obtain an example of the NMSSM, one needs to impose the extra requirement
that there exist at least one  massless singlet which couples to the Higgs pair. 
This condition turns out to be  very restrictive. In particular,  we have analyzed the
``fertile patch'' of the mini--landscape  with \SO{10} local GUTs  of Ref.~\cite{Lebedev:2006kn}  
and found no NMSSM examples. These models
contain 2 Wilson lines and appear to be quite rigid in the sense that the decoupling of
exotics implies  that all the SM singlets are also heavy\footnote{This does not always apply
to models with \E6 local GUTS and we have identified one example leading to  the NMSSM 
in the ``decoupling'' limit \cite{Lebedev:2009xx}.}.   
In setups with 3 Wilson lines of Ref.\cite{Lebedev:2008un}, this is not the case and we have identified
a number of the NMSSM candidates.

\subsection{Example:  NMSSM with approximate Peccei--Quinn symmetry  }

Here we present an example of a string model which matches closely properties of the NMSSM
in the Peccei--Quinn limit.
It is based on a heterotic orbifold \Z6-II. In the notation of Ref.~\cite{Lebedev:2008un}, the model is
defined by the following  gauge shift and Wilson lines (in the E$_8\times$E$_8$ root basis):
\begin{subequations}
\begin{eqnarray}
V & = &\left(\tfrac{1}{6},\,-\tfrac{1}{3},\,-\tfrac{1}{2},\,0,\,0,\,0,\,0,\,0\right)\left(0,\,0,\,0,\,0,\,0,\,0,\,0,\,0\right) 
\;,\\
W_{2} & =
&\left(1,\,\tfrac{1}{2},\,0,\,\tfrac{1}{2},\,\tfrac12,\,-\tfrac12,\,-1,\,0\right)\left(-\tfrac14,\,\tfrac34,\,\tfrac14,\,\tfrac14,\,\tfrac34,\,-\tfrac34,\,-\tfrac34,\,\tfrac34\right)
\;,\\ 
W_{2}' & =
&\left(\tfrac34,\,\tfrac{3}{4},\,-\tfrac{1}{4},-\tfrac{1}{4},\,-\tfrac{1}{4},\,\tfrac34,\,\tfrac14,\,\tfrac14\right)\left(-\tfrac14,\,-\tfrac14,\,-\tfrac14,\,-\tfrac14,\,-\tfrac14,\,\tfrac14,\,\tfrac14,\,\tfrac34\right)
\;,\\ 
W_{3} & = 
&\left(-\tfrac{5}{6},\,-\tfrac{7}{6},\,\tfrac{1}{2},\,\tfrac{1}{2},\,\tfrac{1}{2},\,-\tfrac{1}{2},\,-\tfrac{1}{2},\,-\tfrac{1}{2}\right)\left(0,\,0,\,\tfrac13,\,\tfrac13,\,\tfrac13,\,0,\,1,\,\tfrac23\right)\;.
\end{eqnarray}
\end{subequations}

The gauge group after compactification is
\begin{equation}
G_\mathrm{SM}\times[\SU6]\times\U1^7\;,
\end{equation}
where $G_\mathrm{SM}=\SU3_C\times\SU2_\mathrm{L}\times\U1_Y$ includes the standard \SU5 hypercharge generator
\begin{equation}
\mathsf{t}_Y~=~
\left(0,0,0,\tfrac{1}{3},\tfrac{1}{3},\tfrac{1}{3},-\tfrac{1}{2},-\tfrac{1}{2}\right) \, (0,0,0,0,0,0,0,0)\;.  
\end{equation}
(Here we do not require the existence of non-anomalous $B-L$ symmetry.)
The resulting massless spectrum is displayed in
Table~\ref{tab:spectrum}. At this step, it contains 3 SM families plus
vector--like matter.  One of the SM generations comes from the
$\boldsymbol{27}$--plet of \E6 located at the fixed point at the
origin in the compact space, while the other two come from various
twisted and untwisted sectors. All three generations are intrinsically
different in this model. Further details can be found in~\cite{Lebedev:2009xx}.
 
\begin{table}[!h!]
\begin{center}
\begin{tabular}{|rlc|rlc|c|rlc|}
\cline{1-6}\cline{8-10}
  \#  &  Irrep                                          & Label & \# & Anti-irrep       & Label    &&  \#  & Irrep & Label \\
\cline{1-6}\cline{8-10}
  4 & $( {\bf 3}, {\bf 2};  {\bf 1})_{1/6}$   & $q_i$ &
  1 & $( {\bf\overline{3}}, {\bf 2};  {\bf 1})_{-1/6}$  & $\bar q_i$ &&
 63 & $( {\bf 1}, {\bf 1};  {\bf 1})_{0}$     & $s^0_i$ \\
  9 & $( {\bf 1}, {\bf 2};  {\bf 1})_{-1/2}$  & $\ell_i$ &
  6 & $( {\bf 1}, {\bf 2};  {\bf 1})_{1/2}$      & $\bar\ell_i$ && 
  4 & $( {\bf 1}, {\bf 1};  {\bf\overline{6}})_{0}$    & $\bar h_i$ \\
  4 & $( {\bf\overline{3}}, {\bf 1}; {\bf 1})_{-2/3}$  & $\bar u_i$ &
  1 & $( {\bf 3}, {\bf 1};  {\bf 1})_{2/3}$      & $u_i$ && 
  4 & $( {\bf 1}, {\bf 1};  {\bf 6})_{0}$        & $h_i$ \\ 
\cline{8-10}
  4 & $( {\bf 1}, {\bf 1};  {\bf 1})_{1}$       & $\bar e_i$ &
  1 & $( {\bf 1}, {\bf 1};  {\bf 1})_{-1}$      & $e_i$  &\multicolumn{3}{c}{$\phantom{I^{I^I}}$}\\
  8 & $( {\bf \overline{3}}, {\bf 1};  {\bf 1})_{1/3}$ & $\bar d_i$ &
  5 & $( {\bf 3}, {\bf 1};  {\bf 1})_{-1/3}$     & $d_i$ &\multicolumn{3}{c}{$\phantom{I^{I^I}}$}\\
\cline{1-6}
  1 & $( {\bf 3},  {\bf 1};  {\bf 1})_{1/6}$   & $v_i$ &  
  1 & $( {\bf\overline{3}}, {\bf 1};  {\bf 1})_{-1/6}$  & $\bar v_i$ & \multicolumn{3}{c}{$\phantom{I^{I^I}}$} \\
  1 & $( {\bf 1},  {\bf 1};  {\bf\overline 6})_{1/2}$   & $w^+_i$ &
  1 & $( {\bf 1}, {\bf 1};   {\bf 6})_{-1/2}$ &  $w^-_i$ & \multicolumn{3}{c}{$\phantom{I^{I^I}}$} \\
  9 & $( {\bf 1},  {\bf 1};  {\bf 1})_{1/2}$   & $s^+_i$ &  
  9 & $( {\bf 1},  {\bf 1};  {\bf 1})_{-1/2}$  & $s^-_i$ &  \multicolumn{3}{c}{$\phantom{I^{I^I}}$} \\
  6 & $( {\bf 1},  {\bf 2};  {\bf 1})_{0}$     & $m_i$ &  \multicolumn{3}{c|}{$\phantom{I^{I^I}}$} &  \multicolumn{3}{c}{$\phantom{I^{I^I}}$} \\
\cline{1-6}
\end{tabular}
\caption{Massless spectrum. Representations  with respect to
$[\SU3_C\times\SU2_\mathrm{L}]\times[\SU6]$ are given  in  bold face, the hypercharge is
indicated by the subscript.}
\label{tab:spectrum}
\end{center}
\end{table}

At the next step, many of the SM singlets develop VEVs and break the gauge group to
\begin{equation}
G_\mathrm{SM}\times[\SU6\times\U1]\;,
\end{equation} 
where $[\SU6\times\U1]$ is hidden in the sense that no SM particle is
charged under this group.
At the same time, the unwanted vector--like exotics attain large masses and decouple.
The resulting massless spectrum is that of the MSSM plus, possibly, SM singlets. 

We choose a specific configuration of the SM singlet VEVs, in which only the fields
\begin{equation}
\label{eq:stilde}
 \{\widetilde{s}_i\}
 = \{
s^0_{1}, s^0_{2}, s^0_{7}, s^0_{12}, s^0_{14}, s^0_{21}, s^0_{22}, s^0_{27}, s^0_{30}, s^0_{31},
s^0_{35}, s^0_{37}, s^0_{40}, s^0_{41}, s^0_{44}, s^0_{45}, s^0_{61}, s^0_{63}
 \}  \label{vacuum}
\end{equation}
develop non-zero VEVs, while  
 the expectation values of all other fields vanish.
We have checked that, in this case, the mass matrices for the exotics   have maximal rank.
The cancellation of the Fayet--Iliopoulos term and D--flatness are guaranteed by the 
holomorphic monomial 
\begin{equation}
\label{eq:monom}
\hskip -3mm \psi=
(s^0_{1})^7 (s^0_{2})^5 (s^0_{7})^2 (s^0_{12})^2 (s^0_{14})^4 s^0_{21} s^0_{22} s^0_{27} (s^0_{30})^2 (s^0_{31})^2
s^0_{35} s^0_{37} s^0_{40} s^0_{41} s^0_{44} s^0_{45} (s^0_{61})^6 s^0_{63} \;.
\end{equation}

The massless (to order 6 in singlet VEVs)   pair of Higgses is  given schematically by
\begin{eqnarray} 
&& H_u~\sim~ \bar\ell_1+\widetilde{s}\,\bar\ell_2+\bar\ell_3+\bar\ell_4+\bar\ell_5\;, \nonumber\\
&&  H_d~\sim~\,\ell_9\ \;,
\end{eqnarray}
where $\ell_i, \bar\ell_i$ are  the \SU2 doublets of Table \ref{tab:spectrum} and we have omitted
order 1 coefficients. 
There exists one  massless   SM singlet 
\begin{eqnarray}
  \label{eq:singlet}
 S&=&  s^0_{66}\, 
\end{eqnarray}
and it couples to the Higgs pair at the trilinear order:
\begin{equation}
  \label{eq:muterm}
\lambda ~  S H_u H_d~ = ~ \lambda ~ s^0_{66} \bar\ell_3 \ell_9 ~+~ \text{higher order terms}\,.
\end{equation}
This is a twisted coupling  of the type $T_5 T_5 T_2$, where $T_i$ denotes
a twisted sector.
On the other hand, the self-interaction  $S^3$ is not allowed at least 
to order $\tilde s^5$.
Thus, in terms of Eq.(\ref{nmssm}), we have
\begin{eqnarray}
&& \lambda \sim 1  \;, \nonumber\\
&& \kappa < {\cal O}(\tilde s^5) \;.
\end{eqnarray}
For $\widetilde s <1$, the system has an approximate Peccei--Quinn symmetry,
whose spontaneous breaking results in a light pseudoscalar state $A_{\rm PQ}$.
Its mass depends on the order of the allowed coupling as well as the 
exact value of $\tilde s$  and can be light enough
to be relevant to the MSSM fine-tuning problem. 

Let us also note that the discrete symmetry of this \Z3--NMSSM is
expected to be broken by a small supersymmetric mass term for the
singlet, which helps avoid cosmological problems associated with
spontaneous breaking of discrete symmetries.

The existence of approximate symmetries in the low-energy theory is a result
of the string selection rules combined with a specific choice of the vacuum 
configuration (\ref{vacuum}). From the bottom-up perspective, it is not
transparent why the term $S^3$ for an SM singlet  is not allowed. This becomes
clear in the UV completion of the model: $S$ stems from the \E8$\times$\E8 
sector and thus  carries additional gauge charges.
Similarly, the bare $\mu$--term is not allowed (to order $\tilde s^6$)
due to our choice of the  vacuum configuration  and the string
selection rules for the couplings. Thus, the approximate Peccei-Quinn symmetry
is enforced by  (\ref{vacuum}). Similarly, an example of approximate R--symmetry
was constructed in Ref.~\cite{Kappl:2008ie}.


\section{Conclusion}

We have undertaken a search for the NMSSM in the framework  of the heterotic string
compactified on a \Z6-II orbifold. Although  there are many models with the MSSM
spectrum, the NMSSM--like models are rare. This is due to the additional requirement
that there exist at least one  light SM singlet that couples to the Higgs pair. 
Our search within the ``fertile patch'' of the heterotic landscape  with \SO{10} local GUTs  
\cite{Lebedev:2006kn}  
 has given null results,
yet we have found a number of the NMSSM candidates in setups with 3 Wilson lines \cite{Lebedev:2008un}.

The stringy NMSSMs share the  common feature that the  singlet self-interactions are 
typically suppressed, which leads to specific versions of the NMSSM.  
In one variant, the singlet sector essentially decouples from the MSSM, while in the other
there is an approximate Peccei--Quinn symmetry whose breaking leads to a light pseudoscalar.
The latter can be relevant to the MSSM fine-tuning problem since 
in this case the bound on the Higgs mass  relaxes.

The apparent  (approximate)   
symmetries of the low energy theory result from properties of its stringy
UV completion. For instance, the suppression  of the direct $\mu$-term and the singlet
self--interactions is due to the string selection rules combined with our choice of
the vacuum state. The ensuing Peccei--Quinn symmetry is broken by higher order
terms in the superpotential.

\providecommand{\bysame}{\leavevmode\hbox to3em{\hrulefill}\thinspace}

\end{document}